# MAM-STM: A software for autonomous control of single moieties towards specific surface positions

Bernhard Ramsauer[1], Johannes J. Cartus[1], Oliver T. Hofmann[1*]

1 Institute of Solid State Physics, NAWI Graz, Graz University of Technology, Graz, 8010, Austria

Corresponding author:
o.hofmann@tugraz.at

## ABSTRACT

In this publication we introduce MAM-STM, a software to autonomously manipulate arbitrary moieties towards specific positions on a metal surface utilizing the tip of a scanning tunneling microscope (STM). Finding the optimal manipulation parameters for a specific moiety is challenging and time consuming, even for human experts. MAM-STM combines autonomous data acquisition with a sophisticated Q-learning implementation to determine the optimal bias voltage, the z-approach distance, and the tip position relative to the moiety. This then allows to arrange single molecules and atoms at will. In this work, we provide a tutorial based on a simulated response to offer a comprehensive explanation on how to use and customize MAM-STM. Additionally, we assess the performance of the machine learning algorithm by benchmarking it within a simulated stochastic environment.

## PROGRAM SUMMARY

Program title: MAM-STM

CPC Library link to program files: (to be added by Technical Editor)

Developer's repository link: https://gitlab.tugraz.at/software_public/mam_stm.git

Code Ocean capsule: (to be added by Technical Editor)

Licensing provisions: GNU General Public License 3 (GPL)

Programming language: Python 3

Nature of problem: Achieving precise control over the arrangement of individual molecules on surfaces is essential for advancing nanofabrication and understanding molecular interaction processes. While self-assembly offers a method for forming nanostructures, achieving arbitrary arrangements of moieties remains difficult. Current approaches, such as scanning probe microscopy (SPM), require extensive manual intervention and precise control is difficult to achieve consistently due to the stochastic nature of quantum mechanical systems at the nanoscale. Thus, learning to manipulate several moieties in order to build even relatively small structures is challenging and time-consuming and the automation through conventional expert systems is hindered by the lack of prior knowledge about the surface-moiety interaction processes.

Solution method: This scenario is ideal for machine learning algorithms, like reinforcement learning (RL), which do not require an underlying model but are able to autonomously learn the optimal manipulation parameters by performing manipulations directly at the machine. Introducing MAM-STM, which stands for Molecular and Atomic Manipulation via Scanning Tunneling Microscopy. MAM-STM allows to control molecules and atoms by learning the manipulation parameters for either vertical or lateral manipulations. However, the vast number of manipulation parameter combinations and the inefficient learning procedure of RL agents exhibit several challenges. MAM-STM overcomes these challenges with an autonomous masking routine that eliminates manipulation parameters that induce structural changes to the moiety or lift it off the surface. Additionally, a sophisticated Q-learning approach is developed that speeds up the learning procedure, enabling molecular manipulations within one day of training.

# INTRODUCTION

Achieving atomically precise control over the position and orientation of individual molecules is key for advancing the understanding of crystal growth, assembly processes, and the operation of molecular machines.[1] Such control unleashes the potential for nanofabrication of innovative materials with properties that would be inaccessible through conventional fabrication techniques.[2–6] Self-assembly, driven by lateral intermolecular interactions, represents an established method for forming nanostructures on surfaces.[7,8] Although these structures can be tailored within certain limits through careful selection of functional groups, achieving arbitrary arrangements remains a formidable challenge.[9,10] An alternative, more versatile approach is to form structures building-block by building-block through atomically precise manipulations.[11–13] Currently, this is best achieved using scanning probe microscopy (SPM).

Using this method, it is possible to build artificial structures,[8,14–18] such as quantum corrals,[19–21] or 2D materials.[22,23] Even nano-electronic computational devices, like logic gates,[24–26] can be constructed. However, assembling even moderately sized nanostructures requires hundreds or thousands of manipulation steps with different manipulation parameters for every new type of building block. An additional complication arises from the fact that the interaction processes at the nanoscale are stochastic, i.e. the same manipulation attempt may not always yield the same outcome. This makes finding optimal manipulation parameters repetitive, time consuming, and unintuitive even for experts in the field. Thus, ideally, the process of manipulating surface moieties should be fully automated.

However, conventional expert systems, i.e. algorithms relying on fixed decision processes, require prior knowledge about outcomes — an information which is not available when dealing with moieties on surfaces that are investigated for the first time. Conversely, this is an ideal application for machine-learning algorithms. Beyond mastering highly complex computer games with super-human performance,[27–31] even in the absence of a priori knowledge of game rules, machine learning has found relevance in scientific endeavors. Notably, it has been instrumental in simulating diverse physics experiments,[32–35] enabling autonomous data acquisition in scanning probe microscopy (SPM) experiments,[36,37] and facilitating the detection and movement of nanowires using atomic force microscopy.[38] The ability of scanning tunneling microscopes (STMs) in assembling atoms into atomically perfect nanostructures has been previously demonstrated by using a path planning algorithm and pre-defined manipulation parameters.[14] Furthermore, reinforcement learning (RL) approaches in scanning tunneling microscopy have been employed to remove individual molecules from closed molecular layers,[39] and precisely manipulate single silver atoms to predefined positions on a Ag(111) surface.[40] These advancements collectively underscore the potential of machine learning in enhancing precision and autonomy in the manipulation of atomic and molecular structures. However, a method that learns the manipulation parameters of both atoms as well as molecules in an autonomous fashion does not exist.

In this publication we introduce MAM-STM, a software to control both atoms and molecules towards specific positions on a metal surface by finding the optimal manipulation parameters in an autonomous fashion. MAM-STM stands for _Molecular and Atomic Manipulation via Scanning Tunneling Microscopy_. Fig. 1 exemplifies a typical situation where building blocks on the surface should be arranged into a desired structure, here in the form of an infinity symbol. Fig 1a shows silver adatoms (circled in green) and organic molecules (here: phtalocyanine, $H_2Pc$., circled in red) adsorbed and initially randomly distributed on a silver (111) surface. Furthermore, the blue crosses indicate the planned final location. To apply our method, the first step then is to select a building block and its target location. MAM-STM then automatically moves the moiety to this location by imaging the region around the moiety, applying a vertical or lateral manipulation (see below), and re-imaging a region around the moiety again to evaluate the results. The procedure is repeated until the building block reaches its target location within a pre-defined threshold distance. Figure 1b then shows the state of the surface after manipulating the building blocks using MAM-STM. Our approach requires no prior knowledge of the interaction processes at play i.e., the behavior of the moiety with the tip or the surface, nor the atomistic structure.

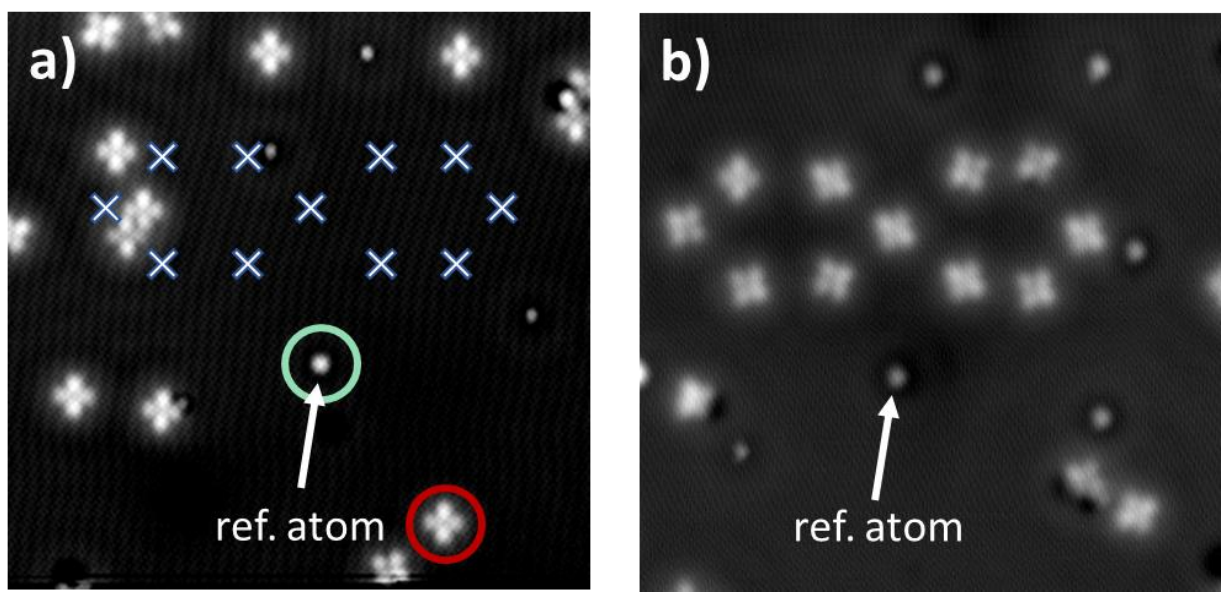

Fig 1a) STM image (U = 0.250 V, I = 100 pA) of the initial state of the surface, containing adatoms (circled in green) and molecules (circled in red). The target position for a functional structure is indicated as blue crosses. b) STM image of the surface (U = 0.250 V, I = 100 pA) where the adsorbates are assembled in the desired structure. The reference atom is indicated by the white arrow.

To obtain the ideal manipulation parameters, the MAM-STM software employs a reinforcement learning approach that adeptly maneuvers individual molecules towards specific positions on a surface using a tip-induced electric field of a scanning tunneling microscope. However, in order to build artificial nanostructure building-block by building-block several challenges have to be addressed. The main challenge is the a priori infinite number of possible manipulations parameter combinations (i.e., the applied bias voltage v, the approach distance z, and the tip position in x and y relative to the moiety), which, in reinforcement learning, is commonly referred to as action space explosion.[41] To overcome this action space explosion, MAM-STM employs an autonomous masking routine that sweeps through the bias voltage and z approach distance to determine an Action Space Mask (ASM) which limits the manipulation parameters to those that are able to induce movements but do not induce any structural changes to the moieties or lift it off the surface. Another challenge is that training conventional reinforcement learning agents is not particularly data efficient. Thus, numerous manipulations have to be carried out necessitating in a substantial amount of measurement time, which is usually limited. MAM-STM employs a distinctive Q-learning approach to enhance the data efficiency of the training process. In this approach, MAM-STM learns from the impact of a single manipulation on all potential surface positions the moiety could attain. This accelerates the learning progress, enabling the precise positioning of moieties after learning a few thousand manipulation parameters.

Following about half a day of training, the algorithm efficiently manipulates initially unknown moieties towards specific positions on the surface. The code of MAM-STM is freely available and can be downloaded from our GitLab repository: **https://gitlab.tugraz.at/software_public/mam_stm.git**

This paper is organized as follows: Firstly, we discuss the general working principle of reinforcement learning and the particular Q-Learning implementation used in MAM-STM. Secondly, we describe how the action space explosion is limited by measuring an Action Space Mask (ASM) to ensure parameters that interrupt the autonomous learning procedure are eliminated. Thirdly, the three key ingredients of reinforcement learning are explained utilizing an artificial molecule as an example. This will give a procedure to follow when setting up MAM-STM for another moiety. Finally, the setup of our reinforcement learning agent is benchmarked by letting the agent learn in a simulated stochastic environment.

## Methods

The software is written in Python3 and the interface to the STM is currently only available for the TMS320C6657 DSP (Digital Signal Processor) from Texas Instruments in order to remote-control the STM. The software was tested on a low-temperature STM (CreaTec) operated at 5 K. However, the code which handles the remote-control commands can be easily adapted to the electronics for other manufactures, like Nanonis.

**Graphical User Interfaces (GUI) for the MAM-STM setup process**

When the software is started, a graphical user interface (GUI), shown in Fig 2, appears which allows to connect the measurements (performed with the commercial CreaTec STM/AFM software) to MAM-STM. The first step is to select a building block (i.e., a starting position) and the position to which it should be moved. To do so, the first step is to obtain an overview image of the surface within the STM/AFM software (using manually chosen and optimized imaging conditions). Then, a build block is selected in the STM/AFM software and read into MAM-STM by clicking the "Read Position" button. The next step is to select the target position in a similar way. If desired, multiple target positions, termed "goal" hereafter, can be selected. This then effectively constitutes the trajectory along which the moiety is manipulated. An example is given in Fig 3, where we selected a molecule as start position and 4 subsequently goals, where the last goal coincides with the starting position leading to a closed trajectory. The last position of the trajectory can be deleted via pressing the "Delete Position" button. A detailed description of the GUI functionality is given in the supporting information.

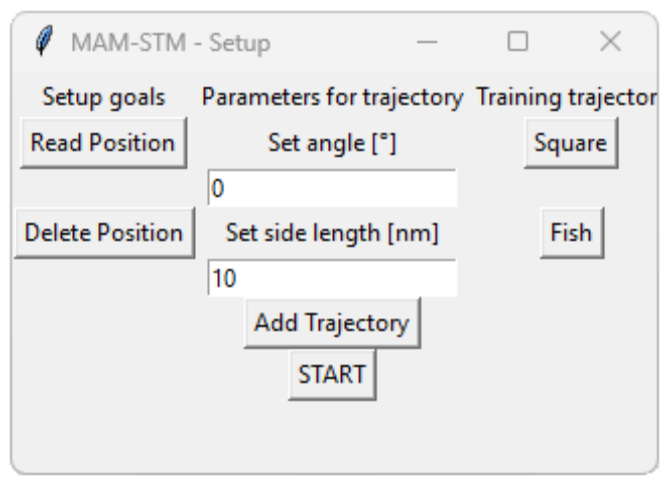

Fig 2 GUI to initialize the environment positions

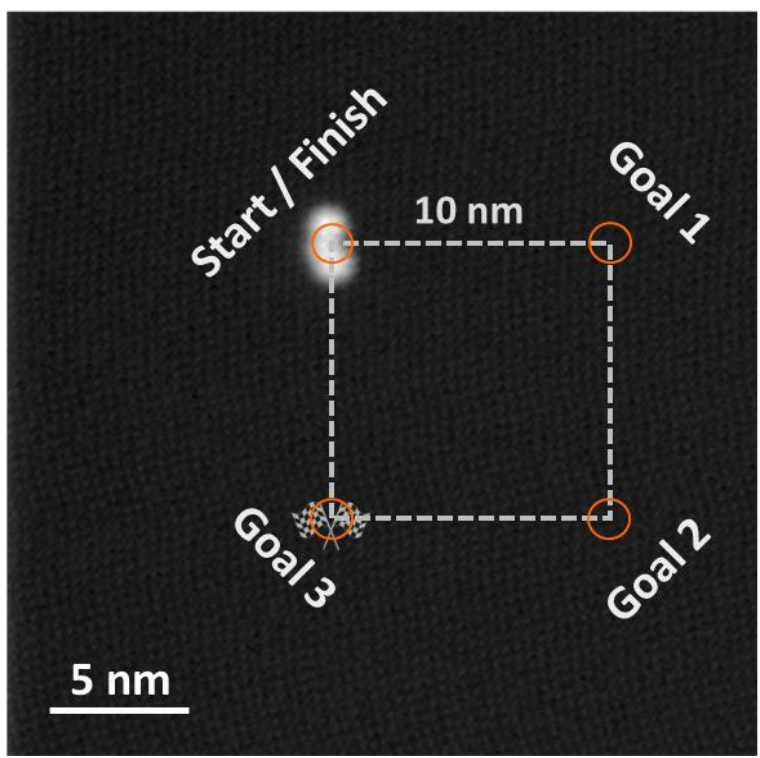

Fig 3 The training trajectory of our RL agent is given by a square of side length 10 nm. The orange circles determined the goal region with a radius of 0.3 nm. The STM-image (1 V, 10 pA) shows the training trajectory of our previous work, [42] where a dipolar molecule is moved across a Ag(111).

# The general working principle of reinforcement learning

Recent advancements have highlighted the efficacy of reinforcement learning techniques in tackling complex and dynamic tasks. In response to these challenges, we have devised a reinforcement learning algorithm tailored to autonomously acquire optimal manipulation parameters. This algorithm controls the STM-tip to directly influence the positioning of moieties on the surface, learning from the outcomes of each manipulation.

1. The machine learning framework

To understand the impact of the various possible settings in this software, it is useful to briefly revisit the foundations of reinforcement learning. Here, we begin by elucidating the general concept of reinforcement learning and then focus on understanding the Bellman equation as it is the central equation (i.e., the Q-learning algorithm) of the learning procedure.[43] This equation is pivotal in determining optimal manipulation parameters for precise control over arbitrary moieties. Additionally, we introduce a strategic approach involving so-called "virtual goals", an innovative trick implemented in MAM-STM to enable parallel learning across all states, thereby enhancing the efficiency of the learning process tremendously. Finally, the specific structure of this reinforcement learning framework is explained, enabling its seamless adaptation to alternative moieties.

*1.1. Reinforcement Learning*

Reinforcement learning (RL) is centered on the concept of learning by applying actions and judging the outcome based on a reward signal, as illustrated in Fig 4.[44] To apply reinforcement learning to precisely position moieties on metal surfaces three key elements must be defined. These are the set of states that represent the goal position relative to the molecule to the agent (shown in Fig 5a), the set of possible actions that the agent can perform (shown in Fig 5b), and the reward function (explained and discussed in Section 1.1.3). At its core, RL addresses the problem of sequential decision-making in dynamic environment, where at each step $t$ an agent interacts with an environment, selects actions, and receives feedback in the form of rewards or penalties.[45] The formal representation is provided through a finite Markov decision process, defined as a 4-tuple $(S, A, P_a, R_a)$.[46] Here, $S$ denotes the set of states $s_t$ referred to as the state space, $A$ represents the set of possible actions $a_t$ collectively denoted as the action space, $P_a(s_t, s_{t+1})$ denotes the state transition probability that when applying action $a_t$ in state $s_t$ the next state $s_{t+1}$ will follow, and $R_a$ signifies the immediate reward $r_t$ received after the state transition. Note that lower-case variables describe a single value of the respective space and upper-case values the complete space.

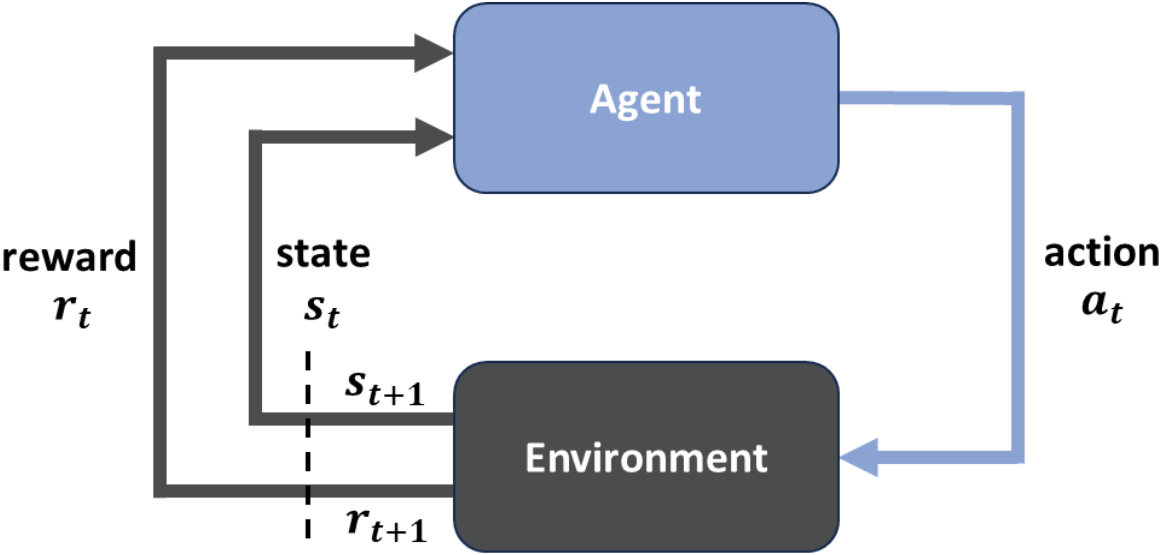

Fig 4 The basic principle of reinforcement learning describes an agent receiving the state $s_t$ from the environment, and it performs an action $a_t$ based on this state. This leads to a transition from state $s_t$ to $s_{t+1}$ and an agent receives the state $s_{t+1}$ and the reward signal $r_{t+1}$.

The primary objective in reinforcement learning is to discover a policy, i.e., a mapping from states to actions, that maximizes the cumulative reward. This policy learning process is guided by the agent's exploration of the environment and its ability to adapt based on the received reward. The iterative nature of RL, involving the agent repeatedly observing, acting, and learning from its experiences, allows it to develop increasingly refined policy $\pi(a|s)$. This policy is determined by a learning algorithm, known as Q-learning (see section 1.2), which maps the action to a particular state via the so-called Q-value $Q(s,a)$. [43]

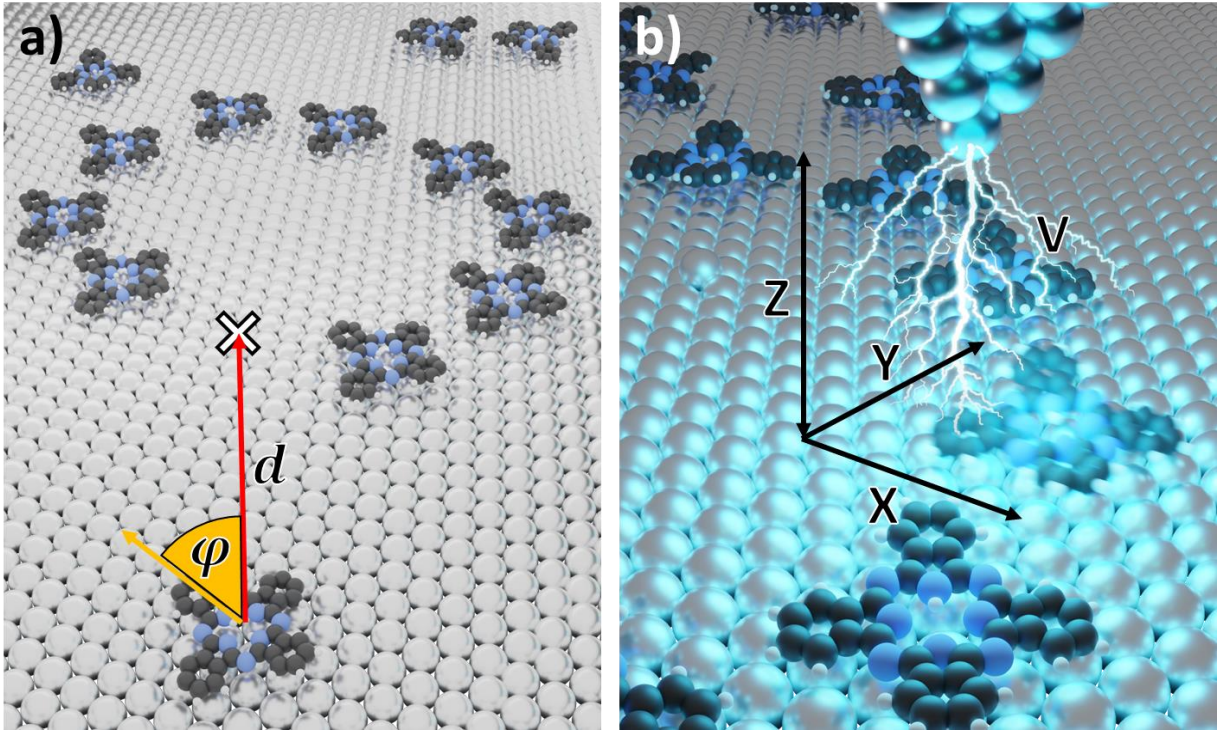

Fig 5 The visual representation of the State-Action Space. a) Each state is given by the angle $\varphi$ and distance $d$ between the goal and the moiety to be moved. The state space encompasses all possible distances and angles b) The action space is defined as the possible parameter combinations given by the bias voltages $V$, approach distance of the tip $Z$, and the relative position of the STM-tip $X, Y$ from the molecules position.

### 1.1.1. State Space

Each individual state is defined as a 2-tuple $(d, \varphi)$, containing the relative angle $\varphi$ between the moiety's orientation and the vector from the moiety position to the goal position, and the distance d the moiety is away from the goal position.

In MAM-STM, the state space $S$ discretized in increments of $\Delta s_\varphi$ for the angle, and $\Delta s_d$ for the distance:

$$S = \{(s_d, s_\varphi) | s_d \in S_d, s_\varphi \in S_\varphi \}$$

with

$$S_d = \{s_d \, | s_d = k \cdot \Delta s_d, k \in \mathbb{N}_0, 0 \leq s_d \leq s_{d_{max}}\}$$

$$S_\varphi = \{s_\varphi \, | s_\varphi = k \cdot \Delta s_\varphi, k \in \mathbb{N}_0, s_{\varphi_{min}} \leq s_\varphi < s_{\varphi_{max}}\}$$

### 1.1.2. Action Space

An action is a combination of manipulation parameters characterized as a 4-tuple $(v, z, y, x)$, where v represents the bias voltage, z denotes the z-approach distance from the STM imaging conditions, and y and x specify the lateral tip position. The action space is also discretized in increments of $\Delta a_v$ for the applied bias voltage, $\Delta a_z$ for the z-approach, $\Delta a_y$ and $\Delta a_x$ for the tip:

$$A_v = \{a_{v_{min}}, a_{v_{min}} + \Delta a_v, a_{v_{min}} + 2\Delta a_v, \ldots, \quad a_{v_{max}}\}$$

$$A_z = \{a_{z_{min}}, a_{z_{min}} + \Delta a_z, a_{z_{min}} + 2\Delta a_z, \ldots, \quad a_{z_{max}}\}$$

$$A_y = \{a_{y_{min}}, a_{y_{min}} + \Delta a_y, a_{y_{min}} + 2\Delta a_y, \ldots, \quad a_{y_{max}}\}$$

$$A_x = \{a_{x_{min}}, a_{x_{min}} + \Delta a_x, a_{x_{min}} + 2\Delta a_x, \ldots, \quad a_{x_{max}}\}$$

$$A = \{(a_v, a_z, a_y, a_x) | a_v \in A_v, a_z \in A_z, a_y \in A_y, a_x \in A_x\}$$

This establishes the set of manipulation parameters available for MAM-STM to utilize in learning the control of individual moieties. Furthermore, MAM-STM can employ both manipulation modes: Vertical and lateral manipulation. Which type of manipulation is executed is selected when calling the program (see Supporting Information for details). As described below, the definition of the lateral tip positions depends on the manipulation mode.

**Pulsing vertical manipulation**

The advantage of the pulsing vertical manipulation mode is that it allows to induce small, but very precise movements. [11,12,42] In the pulsing vertical manipulation mode, the STM is set to the imaging parameters and the tip moves to the agent's selected action $(a_x, a_y)$, shown in red in Figure 5). The center of the action space is defined by the center of the moiety. The size of the action space can be set by a minimum and maximum value in both directions ($a_{x_{min}}$, $a_{x_{max}}$, $a_{y_{min}}$, $a_{y_{max}}$). In practice, it has proven useful to select an action space that is slightly larger than the size of the molecule (see also benchmark below).

$A_x$ and $A_y$ is larger (two times $\Delta l$ or $\Delta w$) than the moiety size (see Fig 6). When the position is reached, the z approach distance is set and the bias voltage is applied for a maximum duration $t_{max}$ or until the current threshold $I_{max}$ (both can be set in the file input.json) is reached.

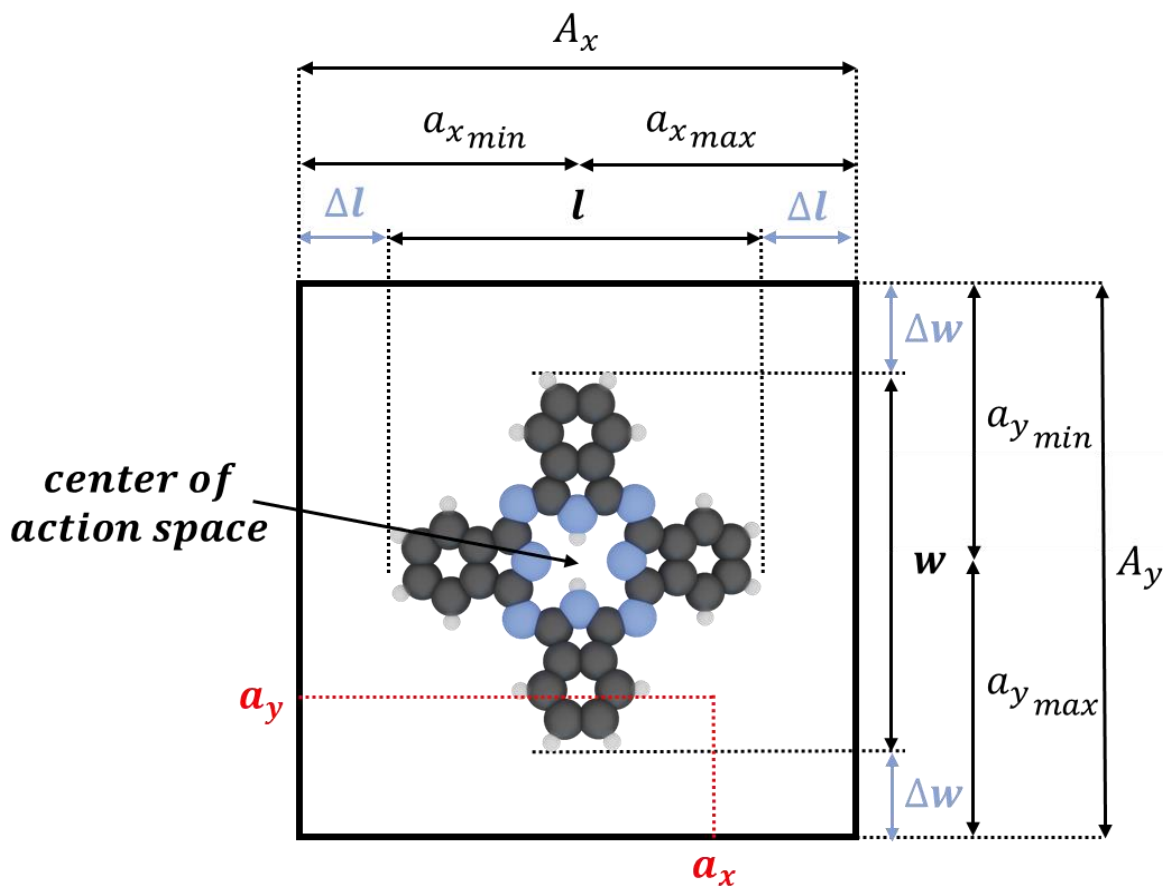

Fig 6 Action space explained based on an exemplary molecule. The center of the action space equals the center of the moiety The action space size $A_x$ and $A_y$ is defined as the minimum and maximum action space distance $a_{x_{min}}$ and $a_{x_{max}}$ or $a_{y_{min}}$ and $a_{y_{max}}$ (i.e., the distance relative to center of the moiety) based on the size of the moiety. These two parameters are useful if the molecule to manipulate is asymmetric. The additional size of the action space $\Delta l$ or $\Delta w$ depends on the type of moiety. In the case of a dipolar molecules is it advisable to use larger values of $\Delta l$ and $\Delta w$ compared to non-polar molecules.

**Lateral manipulation**

Performing a lateral motion of the tip parallel to the surface allows for the moiety to be moved over greater distances compared to the pulsing vertical manipulations. Depending on the manipulation parameters, lateral manipulation leads either to pulling, pushing, or sliding motions of the moiety.[15] The lateral manipulation routine of MAM-STM is depicted in Fig 7. Here, the STM tip moves to the start position of the lateral manipulation (in constant current mode) using the same conditions that are used for imaging. When the start position is reached, the current feedback is turned off and the agent's selected manipulation parameters are used for the manipulation. Thus, the STM tip approaches the surface in z-direction from the imaging condition and applies the bias voltage v.

Then the STM-tip then moves along the lateral manipulation vector towards the end position which is determined by the agent's selected action $(a_x, a_y)$, which can be anywhere within the action space (black square). Contrary to vertical manipulation, in lateral manipulation mode the action space is centered at the goal position. The speed of the lateral manipulation is set, by default, to 2 nm/s and can be adjusted in the input.json file. Note that the lateral manipulation vector starts at an offset $lat_{offset} = \max(l_{molecule}, w_{molecule}) * 1.25$ measured from the moiety center and always goes through the center of the moiety. Thus, the direction of manipulation is given by the end position (i.e., the agent's selected action $(a_x, a_y)$ relative to the goal position).

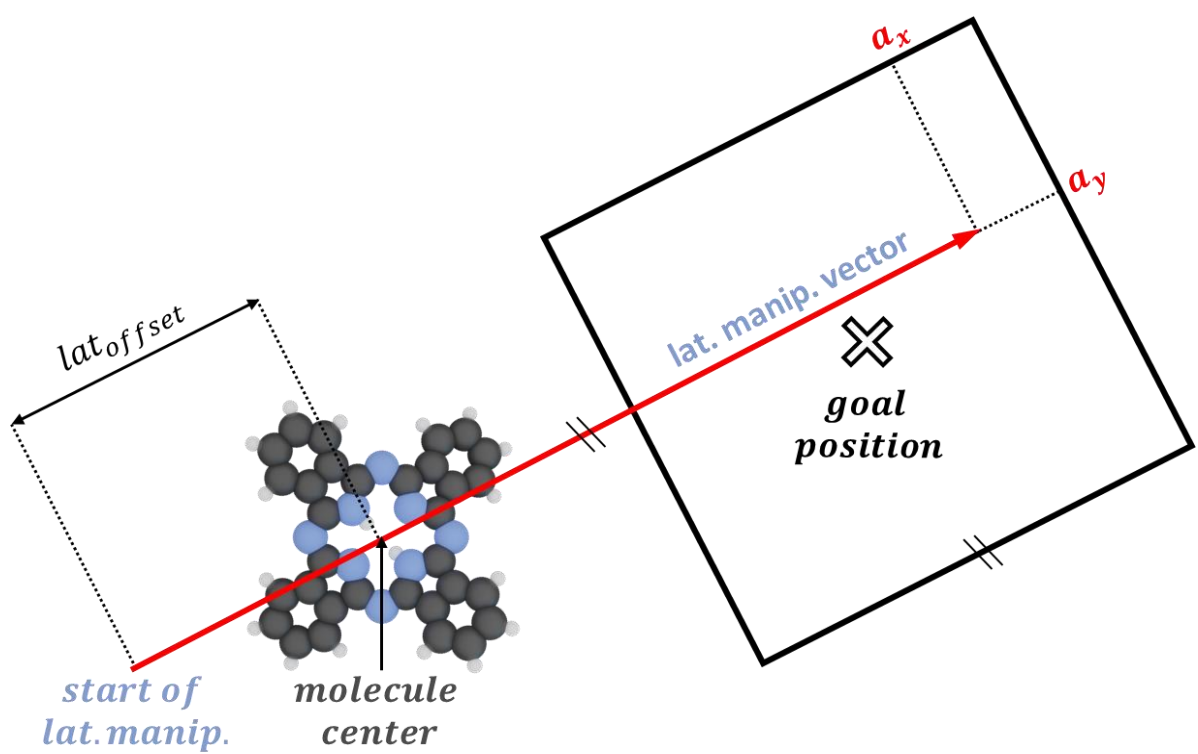

Fig 7 The lateral manipulation across the surface is given by the lateral manipulation vector. The vector starts at a specific distance $lat_{offset} = \max(l_{molecule}, w_{molecule}) * 1.25$ away from the center of the moiety and is displaced along the manipulation vector but in opposite direction of the goal. The end position of the lateral manipulation is determined by the agent's selected action $(a_x, a_y)$ given by the action space (black square) with the action space being centered at the goal position. Note: The shown molecule is only an exemplary molecule used for visualization.

### 1.1.3. Reward functions

In order for the agent to know which action to pick in a given state, it needs to know the quality of an action. This quality is determined via the designed reward function(s) that encodes the overall objective the agent should accomplish. It should be emphasized that this is the only handle we have to influence the agent's behavior. Thus, designing an appropriate reward function is crucial in accomplishing the objective we desire.

In MAM-STM, the agent should achieve two objectives: the first is to manipulate the moiety as fast as possible towards the goal (i.e., in as few steps as possible), and the second objective is to position the moiety precisely at the goal position.

In the following section the reward functions for the two manipulation modes are discussed to understand how we encourage our reinforcement learning agent to manipulate moieties fast and precisely towards specific positions on a metal surface.

**Reward function: vertical manipulation**

In the vertical manipulation the STM-tip is placed based on the selected action $(a_x, a_y)$. The movement caused by this action is converted into a reward signal for the agent. The moiety is controlled towards the individual (sub-)goal positions by applying consecutive voltage pulses until the final goal is reached.

To ensure the agent learns to move the moiety over large distances when being further away from the goal position but precisely when being close to the goal position, two reward functions are designed and then added together. The first dominates when the moiety is far away from the goal position. It is a sigmoid (see Fig 8a) that gives positive reward $r_{t+1}$ when moving towards the goal ($\Delta d_{goal} > 0$) and negative reward when the moiety moves away from the goal position. This incentivizes the agent to move the moiety a large distance $\Delta d_{goal}$ towards the goal. The second reward function (Fig 8b) dominates when the moiety is close to the goal and encourages the agent to move the moiety precisely at the goal position. The y-axis is normalized by $a_{max} = \max(a_{x_{min}}, a_{x_{max}}, a_{y_{min}}, a_{y_{max}})$, the maximum displacement of the STM-tip in x or y relative to the moiety center.

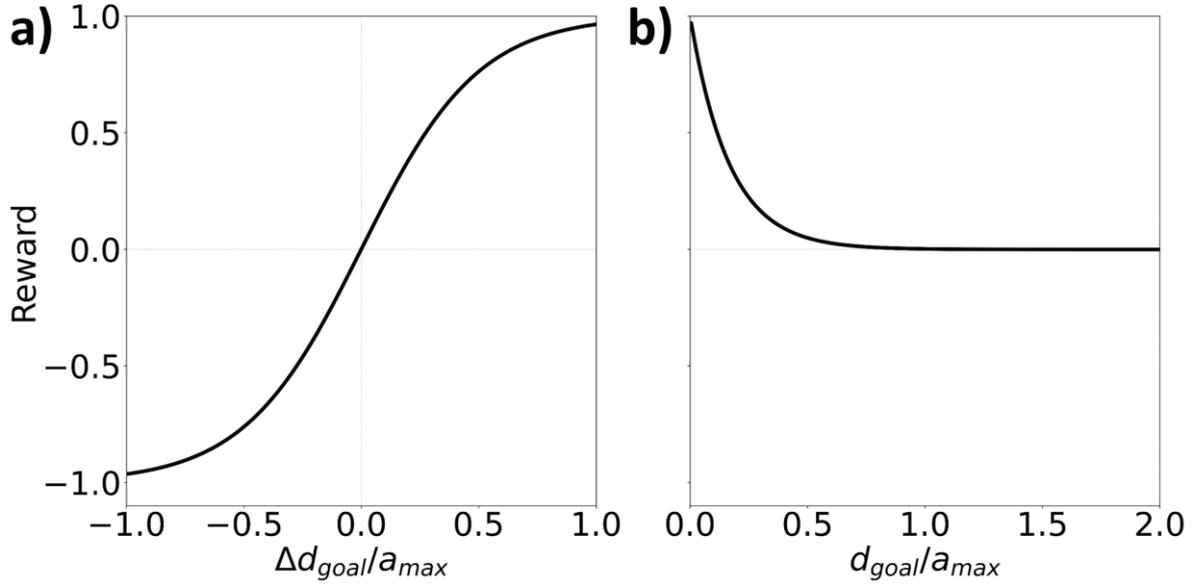

Fig 8 The designed reward functions for the vertical manipulation. a) The reward signal R of the sigmoid function encourages the agent to move the moieties over large distances Δd when the moiety is further away from the goal. The x-axis is normalized by $a_{max} = \max(a_{x_{min}}, a_{x_{max}}, a_{y_{min}}, a_{y_{max}})$ the maximum displacement of the STM-tip in x or y relative to the moiety center. b) This reward function enables additional reward when the moiety is positioned precisely at the goal position and only gives a positive contribution when the moiety is in close proximity to the goal. The x-axis is determined by the distance the moiety is away from the goal $d_{goal}$ normalized by the maximum size of the action space in x and y.

**Reward function: lateral manipulation**

The designed reward function for lateral manipulation (shown Fig 9) closely mirrors the reward function employed for vertical manipulation (refer to Fig 8b), with the inclusion of an additional penalty term, as given in the Supporting Information. This supplementary term is incorporated to penalize instances where the moiety is not precisely situated at the goal position. This is required to penalize situation where the moiety becomes "dropped" during the manipulation, e.g. because of sub-optimal manipulation parameters. The penalty is structured as a linear function, with the slope configured in a manner that imposes a penalty of -0.25 when the distance of the moiety to the goal is $a_{max} = \max(a_{x_{min}}, a_{x_{max}}, a_{y_{min}}, a_{y_{max}})$. This reward $r_{t+1}$ encourages the agent to seek manipulation parameters that accurately position the moiety at the specified goal, discouraging parameters that result in imprecise positioning or potential loss of the moiety throughout the manipulation trajectory.

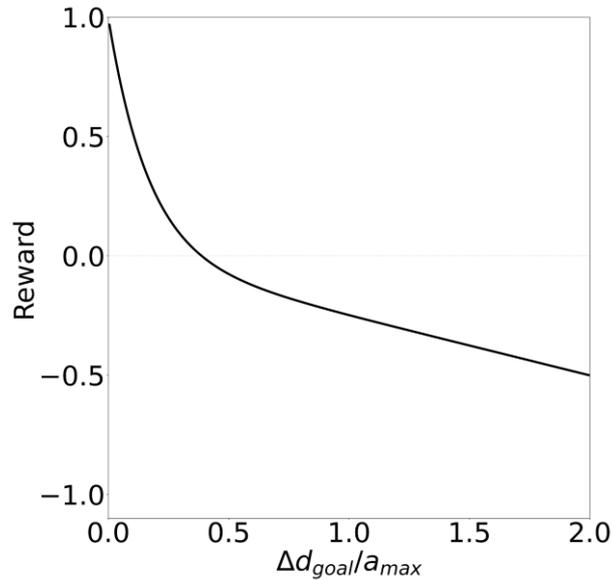

Fig 9 The designed reward function for the lateral manipulation. The reward function encourages the agent to move the moiety exactly to the goal position. The linear penalty function (of slope -0.25 and offset 0) adds a higher penalty the further the moiety is away from the goal position. The penalty term is defined such that a penalty of -0.25 is given if the moiety is positioned at $a_{max} = \max(a_{x_{min}}, a_{x_{max}}, a_{y_{min}}, a_{y_{max}})$.

In summary, we defined three key elements: the states perceivable by the agent, the array of actions it can choose to manipulate the moiety, and the reward function(s) that judge the quality of an action. These components collectively lay the foundation for configuring a reinforcement learning agent capable of directing moieties to predetermined positions on a surface. However, the expansive nature of the state-action space poses several challenges which on the one hand necessitates in eliminating the state dependence in the learning procedure and on the other hand strategically truncate of the action space by determining an autonomous action space mask that ensures only sensible set of parameters are used during training consequently enhancing the efficiency of the learning procedure.

*1.2. Q-learning algorithm*

At the heart of the Q-Learning algorithm is the Q-table, which is a data structure used to store and update the estimated Q-values for the expected cumulative rewards of state-action pairs. The Q-table is initialized with zeros, and Q-Learning iteratively refines these values by applying the Bellman equation.

**Bellman Equation**

The Bellman equation expresses the relationship between the Q-value in the current state and the Q-values of the next state. It enables Q-Learning to update the Q-values through the following formula:

$$Q(s_t, a_t) \leftarrow Q(s_t, a_t) + \alpha \left[ r_{t+1} + \gamma \max_{a'} Q(s_{t+1}, a') - Q(s_t, a_t) \right] \qquad (1)$$

Here, $Q(s_t, a_t)$ is the Q-value for state $s_t$ and action $a_t$, $R_{t+1}$ is the immediate reward (explained in the previous section 1.1.3), $\alpha$ is the learning rate, $\gamma$ is the discount factor.

The learning rate $\alpha$ controls the step size in updating the Q-values. It determines how much the algorithm trusts newly acquired information. A high learning rate may lead to faster convergence but risks overshooting optimal values, while a low learning rate ensures stability but slower learning. Finding an appropriate learning rate is often an empirical challenge in Q-Learning.

The discount factor $\gamma$ allows the RL agent to balance the trade-off between immediate and future rewards. It influences the agent's decision-making process by affecting its preference for short-term gains versus long-term objectives.

During the learning process, the agent balances exploration (trying hitherto unvisited actions) and exploitation (apply the currently action estimated to yield the best result and confirming the efficiency).

In Q-Learning agent's employ an epsilon-greedy strategy given by:

$$\pi(a|s) = \begin{cases} 1 - \epsilon + \frac{\epsilon}{|A(\epsilon)|} & \text{if } a = \mathrm{argmax}(Q(s,a)), \\ \frac{\epsilon}{|A(\epsilon)|} & \text{otherwise} \end{cases} \qquad (2)$$

where, with probability $\epsilon \in (0,1]$, the agent explores by selecting a random action, and with probability $1 - \epsilon$, it exploits by choosing the action with the highest Q-value. The choice of $\epsilon$ is a critical hyperparameter influencing the algorithm's performance and can be set in the input.json file.

Nevertheless, employing the Bellman equation in this manner only updates a single Q-value corresponding to the action taken in a specific state. This means the agent to manipulate several million parameters to explore each action across all states at least once. This makes the identification of meaningful actions (i.e., manipulation parameters) impractical, prompting us to devise a solution to overcome this challenge.

*1.3. Parallel Learning*

Consequently, our modification to the conventional Q-learning algorithm introduces a distinctive approach, where Q-values for every state are learned simultaneously, removing the necessity to explore actions for every state separately.

This method incorporates virtual goals to update the Q-value for a specific action in every state at the same time. The real state defines the actual goal position relative to the present position of the moiety. The chosen action of the agent during the learning phase are determined based on the real state. However, once an action is performed, its outcome can also be evaluated assuming the goal would have been somewhere else. (In that case, the action was probably sub-optimal).

Consequently, rather than updating the Q-value only for the real state, we can simultaneously update the Q-value of all states. This ensures that all states receive updates at every step, establishing a learning progress that quickly converges independently of the number of states.

*1.4. Action Space Masking: Truncating the Manipulation Parameter Space*

The challenge of the large number of possible manipulation parameters is often referred to as the action space explosion in reinforcement learning. MAM-STM addresses this challenge with a crucial software component called Action Space Masking (ASM). This routine plays a pivotal role in limiting the manipulation parameters, effectively serving as a filter for optimal actions. The importance of the ASM lies in its ability to sweep through the bias voltages and z-approach distances, determining a restricted set of manipulation parameters. By doing so, the ASM ensures that only those parameters capable of inducing movements without causing structural changes are retained. This strategic limitation is essential for preventing the inadvertent imposition of strong manipulation parameters that could lead to undesirable outcomes, such as lifting the moiety off the surface. In essence, the ASM is a key mechanism for refining and streamlining the learning process, enhancing both the precision and safety of the autonomous molecular manipulation conducted by MAM-STM.

The ASM is measured by repeatedly performing lateral manipulations over a specific distance (default: 5 nm), maneuvering the moiety between two fixed points which are determined at the start of the routine. The start fixed point $\vec{v}_{fix\_start}$ is determined by the position of the moiety and the end fixed point is given by a fixed lateral manipulation distance $d_{lat}$ (depends on the moiety size) along the moiety orientation $\varphi_{moiety}$.

$$\vec{v}_{fix\_start} = \vec{v}_{moiety} \tag{3}$$

$$\vec{v}_{fix\_end} = \vec{v}_{moiety} + d_{lat} * \hat{R}(\varphi_{moiety}) \tag{4}$$

with $\hat{R}$ being a rotation matrix.

In general, the manipulation is always along equivalent directions of the moiety, which means every lateral manipulation is either performed in the direction of the moiety or in its reverse direction. To determine which direction of manipulation is performed, the moiety distance towards the two fixed points is determined. The fixed point ($\vec{v}_{fix\_start}$ or $\vec{v}_{fix\_end}$) that is positioned further away from the moiety is the current reference point $\vec{v}_{ref}$ and the corresponding end position of the lateral manipulation $\vec{v}_{lat\_end}$ is determined as follows:

$$\vec{v}_{lat\_end} = \vec{v}_{moiety} + d_{lat} * \hat{R}(\varphi_{lat} + \theta) \tag{5}$$

where $\theta$ is either set to be 0 ° or 180 ° depending on which angle minimizes the distance $\Delta d$ between $\vec{v}_{lat\_end}$ and $\vec{v}_{ref}$:

$$\Delta d = \|\vec{v}_{lat\_end} - \vec{v}_{ref}\| \tag{6}$$

This lateral manipulation routine maneuvers the moiety for a fixed distance $d_{lat}$ back and forth along the moiety's orientation direction $\hat{R}(\varphi_{moiety})$ while also staying as close to the two fixed points as possible.

This preselection routine narrows down bias voltages and approach distances to avoid ineffective or disruptive manipulation, establishing an Action Space Mask to retain suitable parameters during manipulations. An experimentally measured ASM of phthalocyanine on Ag(111) is shown in Fig 10. The colored dots indicate whether the molecule got picked up (grey), moved (pink), or stayed the same (teal) for the given bias voltage v and approach distance z. The ASM shows that as the approach distance increases the necessary bias voltage decreases in order to induce motion. At 0.4 nm approach distance, every bias voltage is able to induce motion, indicating the STM tip being in contact with the molecule. However, increasing the approach distance further causes the molecule to be picked up.

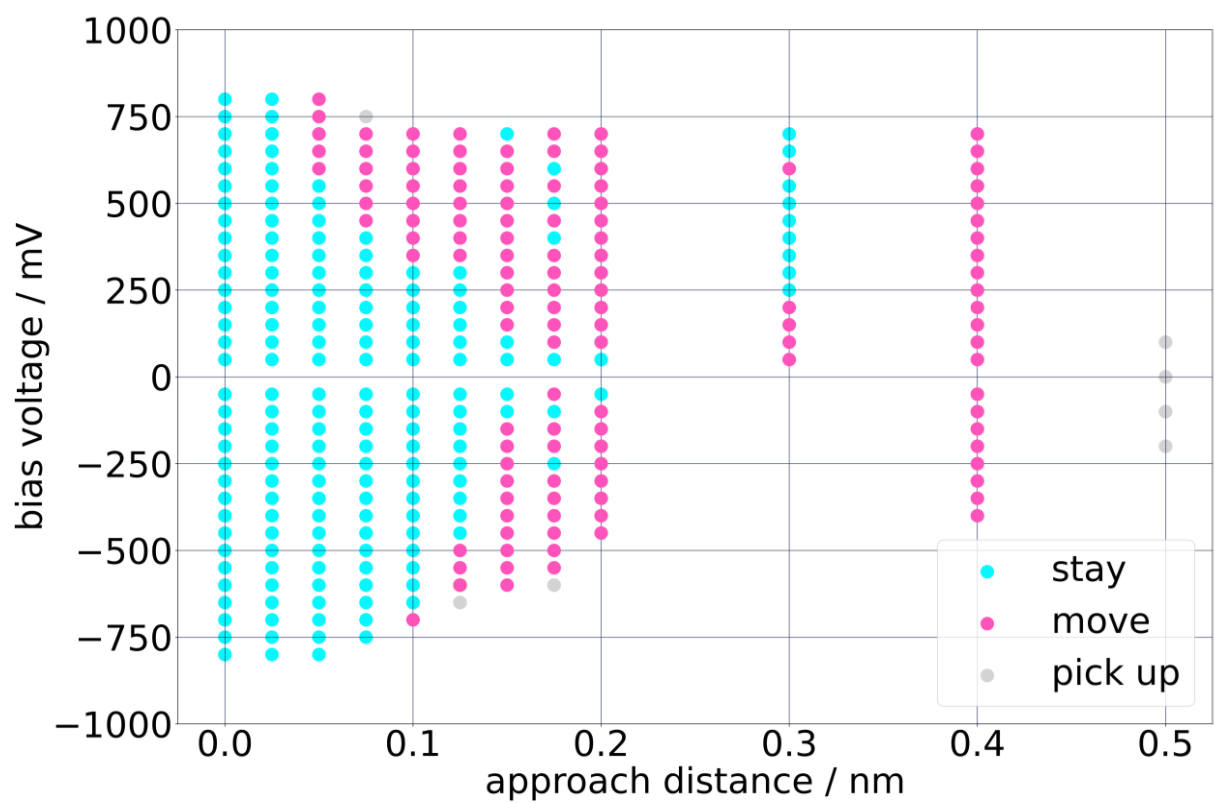

Fig 10 The Action Space Mask of phthalocyanine on Ag(111). The dots represent a single lateral manipulation performed with a specific bias voltage and approach distance. The color indicates whether the molecule stays at the same position, moves, or gets picked up due to the lateral manipulation.

*1.5. Brief description of the code*

The flowchart visualized in Fig 11 gives an overview of the code structure. At the start the individual moieties and their sub- and final goal positions are setup using the *Createc STM/AFM software* in conjunction with the GUI (see Methods). The initialization of the reinforcement learning parameters are loaded from the input.json file. In a real experiment the Action Space Mask (ASM) is measured before the agent is trained on a specific moiety. Once the ASM is measured the RL agent is trained with the truncated action space by sequentially maneuvering every initialized moiety through its defined (sub)goal positions. In order to learn in a continuous fashion a closed trajectory is preferred, thus, the start and final goal position are set to be equivalent.

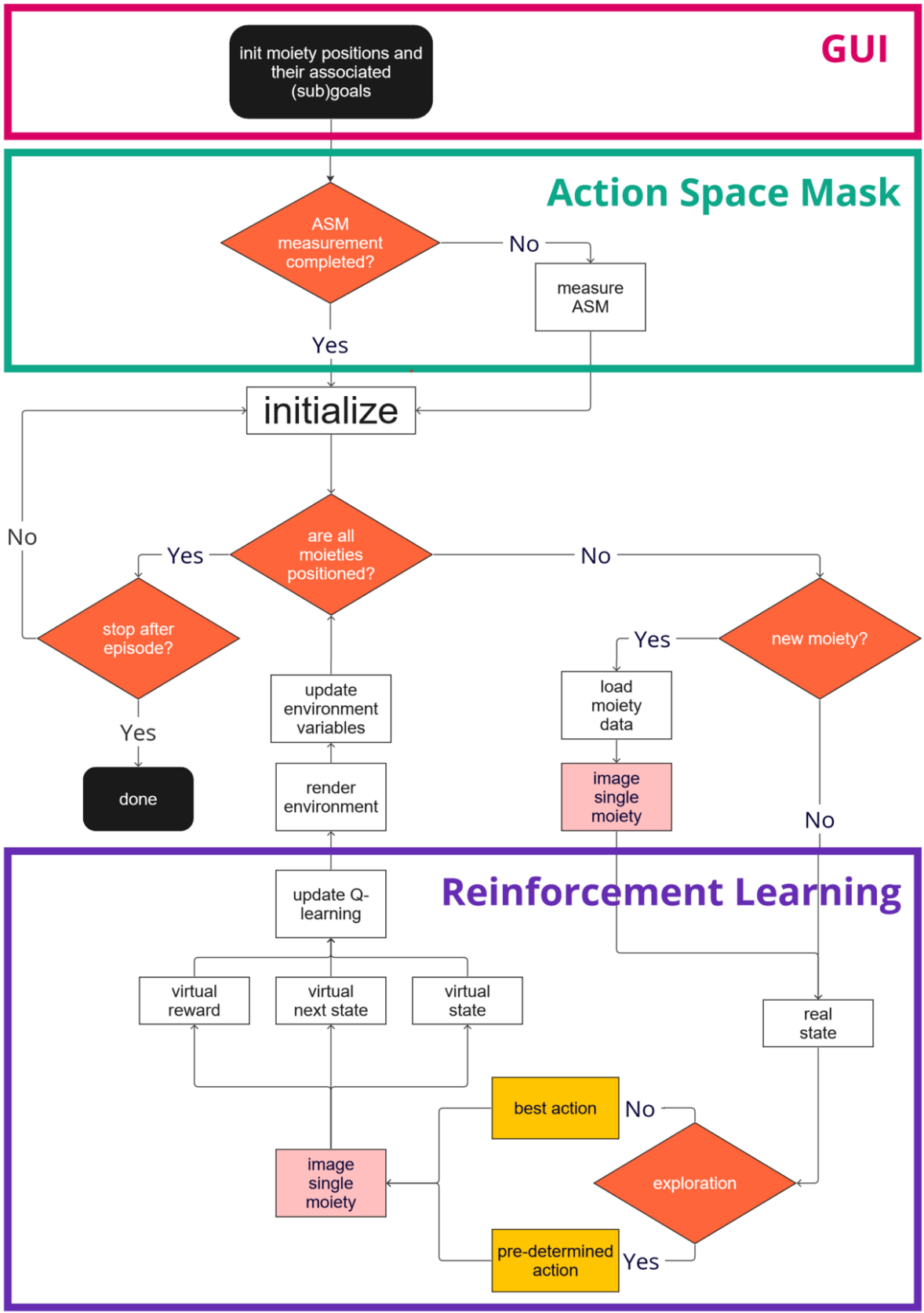

Fig 11 Flow diagram of the MAM-STM software

*1.6. The input.json file*

The complete setup can be configured in the ***input.json*** file located in the MAM-STM's root directory. The functionality of each parameter is meticulously described in the file itself.

# Benchmarking the performance of the agent

*2.1. Simulated responses*

Since measurement time at STMs is not always readily available and the optimization of the learning procedure can be tedious, we implemented a simulation in MAM-STM which allows to consider a hypothetical moiety with pre-defined, realistic and stochastic responses to actions. Within this simulation, one can create an artificial environment, where each manipulation parameter influences the motion of the moiety based on a pre-determined function. For the sake of this benchmark, we choose an environment which mimics the complexity of molecular motion on surfaces. For each manipulation parameter of an action the molecular translation and rotation is affected in a different manner. First the translation and then the rotation is described.

**Translation:** The bias voltage enables the agent to either pull or push the moiety when the bias voltage v is positive or negative, respectively. Furthermore, the absolute value of the bias voltage also allows to increase or decrease the translation distance. The approach distance z can further increase the translation distance by an additional factor.

The distance in x and y the tip is placed relative to the center of the moiety exhibits one point in each quadrant $x_{fix}$ and $y_{fix}$ and which has a fixed translation probability of 100 %. The translation probability decreases according to a Gaussian distribution. Thus, only points in the vicinity of these 4 points are able to induce motion. In addition, to account for the stochastic nature of molecules on surfaces, a Gaussian noise term $G_{noise}(z)$ is added to the motion of the moiety. However, this term only comes into play if the manipulation parameters chosen by the agent are able to induce movement at all. The movement from the old position $x_t$ and $y_t$ to the new position $x_{t+1}$ and $y_{t+1}$ is described by the following equations:

$$x_{t+1} = x_t + G_{a_x}(a_x) + a_x * a_v + sign(a_v) * a_h + G_{noise}(v),$$

$$y_{t+1} = y_t + G_{a_y}(a_y) + a_y * a_v + sign(a_v) * a_h + G_{noise}(v),$$

where $G_{a_x}(a_x)$ and $G_{a_y}(a_y)$ being defined via four Gaussian distributions given by:

$$G_{a_x}(a_x) = \frac{1}{\sigma\sqrt{2\pi}} \left( \sum_{i=1}^{4} e^{-\frac{(a_x-x_{fix,i})^2}{2\sigma_{fix}^2}} \right),$$

$$G_{a_y}(a_y) = \frac{1}{\sigma\sqrt{2\pi}} \left( \sum_{i=1}^{4} e^{-\frac{(a_y-y_{fix,i})^2}{2\sigma_{fix}^2}} \right)$$

where the parametrized Gaussian is defined by $\sigma_{fix} = 0.50\ nm$ and the fixed points in each quadrant:

$$\vec{x}_{fix} = (0.50, -0.40, 0.50, -0.50)^T$$

$$\vec{y}_{fix} = (0.50, -0.75, -0.75, 0.50)^T$$

and the Gaussian noise being:

$$G_{noise}(v) = \frac{1}{\sigma\sqrt{2\pi}} e^{-\frac{(v-\mu)^2}{2\sigma^2}},$$

where the parametrized Gaussian is defined by $\mu = 0$ and $\sigma = 0.17\ nm$.

**Rotation:** The molecular rotation only depends on the lateral position in y relative to the molecule orientation $\varphi_t$. The molecule possesses a distinct internal orientation, subject to alteration when the tip is positioned to the right or left, causing the molecule to rotate either clockwise or counterclockwise by a discrete angle of 60 °. The rotation of the moiety is defined as follows:

$$\varphi_{t+1} = \varphi_t + \text{sign}(a_y) * \frac{\pi}{3} * G_{rot}(v),$$

and follows a Gaussian distribution:

$$G_{rot}(v) = \frac{1}{\sigma\sqrt{2\pi}} e^{-\frac{(z-\mu)^2}{2\sigma^2}}$$

with the gaussian being parametrized by $\mu = 0$ and $\sigma = 1$.

## 2.2. Vertical manipulation benchmarks

The performed benchmark utilizes vertical manipulations for rigorous testing, as it provides a more challenging scenario than the lateral manipulation. Thus, yield to deeper insights into the learning performance of the agent.

The benchmark is performed using MAM-STM's built-in simulation, which allows us to test the reinforcement learning setup and evaluate the performance of an agent in a self-designed artificial environment. In this environment, each manipulation parameter influences the motion of the moiety based on a pre-determined function. However, to account for the stochastic nature of real experiments we added statistical fluctuations in output quantities (i.e., the movement and rotation of the moiety). Thus, we have full control over the effect of the individual manipulation parameters at work, which allows us to benchmark the behavior of the reinforcement learning agent.

Given that time is not crucial and simulating an ASM routine does not provide any benefit, the Action Space Mask (explained in section 1.4) is not measured and we assume valid bias voltages v and approach distances z are already determined. The state and action space used in this benchmark are given in Table 1 and Table 2, respectively, and its values can be adjusted in the input.json file.

Table 1: The state space values for benchmarking

| | Environment |
|---|---|
| | *State Space* |
| Relative distance to goal $d$ | $S_{d\,min} = 0\,nm$ <br> $S_{v\,max} = 5.00\,nm$ <br> $\Delta S_d = 0.20\,nm$ |
| Relative angle to goal $\varphi$ | $S_{\varphi\,min} = -180\,°$ <br> $S_{\varphi\,max} = 180\,°$ <br> $\Delta S_\varphi = 1\,°$ |

Table 2: The action space values for benchmarking

| | Action Space |
|---|---|
| Bias Voltage $V$ | $A_{v\,min} = -800\,mV$ <br> $A_{v\,max} = 800\,mV$ <br> $\Delta A_v = 100\,mV$ |
| Approach distance $Z$ | $A_{z\,min} = 0\,nm$ <br> $A_{z\,max} = 0.50\,nm$ <br> $\Delta A_z = 0.05\,nm$ |
| Relative tip position $X, Y$ | $A_{x\,min} = A_{y\,min} = -1.25\,nm$ |

| | $A_{x_{max}} = A_{y_{max}} = 1.25\ nm$ |
| --- | --- |
| | $\Delta A_y = 0.25\ nm$ |

## *2.3. Training process*

The agent is trained by maneuvering the moiety along a pre-defined trajectory. In the example benchmark we aim for a 10 nm by 10 nm square trajectory, see Fig 3. In the training process, the moiety should consecutively reach the four individual (sub)goal positions of the squared training trajectory. A goal is reached if the moiety is within the threshold distance (orange circle). Note, the goal region (i.e., a threshold distances) can be set in the input.json file.

The agent performs vertical manipulations to maneuver the moiety along the square trajectory and learns the optimal manipulation parameters by maximizing the expected future reward. The reward the agent receives per timestep is given by reward functions elaborated in section 1.1.3. This training trajectory allows the agent to learn continuously and without being interrupted. The agent's hyperparameters during the training procedure are given in Table 3.

Table 3: The Q-learning hyperparameters.

| **Q-learning** | |
| --- | --- |
| *Hyperparameters* | |
| Exploration rate $\epsilon$ | 0.30 |
| Learning rate $\alpha$ | 0.001 |
| Discount factor $\gamma$ | 0.50 |

The discount factor $\gamma$ was chosen rather small (typically around 0.95) so that the agent values the current reward more than the future reward given by the maximum Q-value of the next state. In practice, the shape of the STM tip can change, leading to a different behavior. Thus, with a lower discount factor the agent focuses more on the present than on past experiences.

## 3. Results

The agent is benchmarked by performing validation runs to determine a learning curve. This learning curve is given by the number of manipulations the agent requires to move the molecule along the training trajectory (square), recorded at various stages of training. During the validation runs the agent's knowledge is "frozen" (i.e., the agent is not learning) while moving the moiety along the trajectory.

*3.1. Learning curve*

The learning curve (see Fig 12), shows the agent's performance by traversing the moiety through the four goal positions that form the square trajectory. A single point in the learning curve is determined by performing 10 validation runs at a after a given number of steps for the learning process and determining the mean number of manipulations required to complete the square trajectory. The learning progression is defined as the number of new manipulation parameter combinations the agent has explored. The validation runs are performed after every 100 learned steps. In this benchmarking example, the total number of manipulation parameter combinations is 4235 due to the size of the action space.

In the initial training phase, i.e., from 0 to 400 manipulations (i.e., about 10 % of the total action space), the relative performance of the agent increases by 59 % as the agent captures the coarse dynamics of the environment. In the fine-tuning phase, i.e., from 400 to 4000, the agent's relative performance increases by another 38 %. This increase comes from the agent learning to position the moiety precisely at the goal position, whereas in the beginning it learned to generally move the moiety.

The learning curve decays exponentially, and the agent required only 400 learning steps to find a solid repertoire of manipulation parameters to maneuver an initially unknown moiety towards the individual (sub)goal positions.

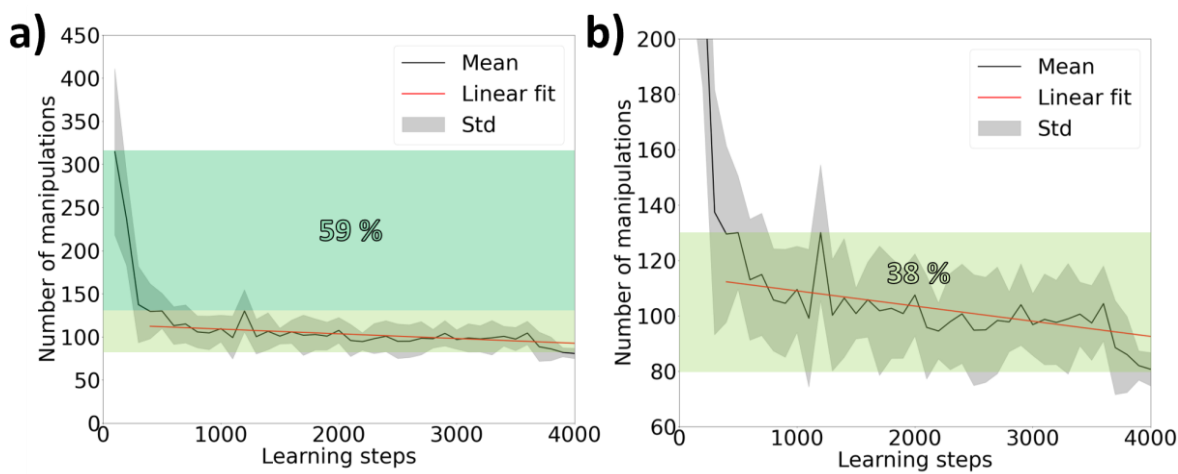

Fig 12 Learning curve of the agent shows the average number of manipulations it requires to move the moiety along the squared training trajectory for a given learning progression. The learning progression is the amount of unique new manipulation parameters the agent has learned. Within the first 400 manipulations, the agent's relative performance increased by 59 % and the remaining learning increases the relative performance by another 38 %.

While the demonstration of MAM-STM here relies on a simulation, we emphasize that it can be readily applied to real experiments. In a previous work, we showed how MAM-STM learns the optimal manipulation parameters for a dipolar molecule on a Ag(111) surface.[42] The RL agent trained for 2250 manipulations to learn where the STM tip has to be placed in order to move the molecule with as few manipulations as possible towards the individual goal positions. Our agent was able to move the molecule with a probability of 82 % towards the individual goal position.

# Conclusion

In this publication we present MAM-STM, a comprehensive tool for learning the manipulation parameters of arbitrary moieties on a metal surface using the tip of a scanning tunneling microscope. MAM-STM is able to learn the manipulation parameters by applying reinforcement learning based on a unique Q-learning approach that eliminates the number of states that have to be learned by the introduction of virtual goals (i.e., all possible goal positions a moiety can end up on the surface). The action space explosion (i.e., the millions of possible manipulation parameter combinations) can be eliminated by measuring an Action Space Mask (ASM) in an automated fashion. In a real experiment, this allows us to eliminate harsh parameters that would hinder learning by inducing structural changes to the moiety or lifting it off the surface, consequently altering the STM-tip. In order to ensure the reinforcement learning framework is set up correctly, MAM-STM's built-in simulator can be used to train an agent in a self-designed artificial environment, allowing to identify problems in advance.

Using MAM-STM's built-in simulation suite, we demonstrated that the agent learned to control an artificial moiety in a self-designed complex environment and its performance was evaluated for about 4000 newly learned manipulations by measuring the learning curve. The learning curve showed that in the first 10% of the newly learned manipulations, the agent's performance increased by 59% and the remaining newly learned manipulations its performance increased by another 38%. This shows that the reward functions as well as the designed state-action space are suitable for solving stochastic environments, like it is the case for real experiments where arbitrary moieties are positioned precisely on a metal surface.

## Supporting information

Getting started, Step-by-Step walkthrough of the setup procedure and how to run MAM-STM, Formal description of the reward functions

## Acknowledgement

We thank Gerhard Meyer (CreaTec) for valuable help with the measurement software of the low temperature STM. We thank Christophe Nacci for careful preparation of the samples as well as Grant J. Simpson and Leonhard Grill (all from University of Graz) for experimental support during the manipulation experiments. We also want to thank, Simon Hollweger, Richard Berger, Lukas Hörmann, and Christoph Wachter for their valuable contributions. B.R., J.J.C., and O.T.H. gratefully acknowledge the funding through the Austrian Science Fund (FWF) (project MAP-DESIGN No.: Y1157-N36).

## REFERENCES

(1)  Simpson, G. J.; Persson, M.; Grill, L. Adsorbate Motors for Unidirectional Translation and Transport. *Nature 2023 621:7977* **2023**, *621* (7977), 82–86. https://doi.org/10.1038/s41586-023-06384-y.

(2)  Civita, D.; Kolmer, M.; Simpson, G. J.; Li, A. P.; Hecht, S.; Grill, L. Control of Long-Distance Motion of Single Molecules on a Surface. *Science (1979)* **2020**, *370* (6519), 957–960. https://doi.org/10.1126/science.abd0696.

(3)  Lastapis, M.; Martin, M.; Riedel, D.; Hellner, L.; Comtet, G.; Dujardin, G. Picometer-Scale Electronic Control of Molecular Dynamics inside a Single Molecule. *Science (1979)* **2005**, *308* (5724), 1000–1003. https://doi.org/10.1126/science.1108048.

(4) Lafferentz, L.; Ample, F.; Yu, H.; Hecht, S.; Joachim, C.; Grill, L. Conductance of a Single Conjugated Polymer as a Continuous Function of Its Length. *Science (1979)* **2009**, *323* (5918), 1193–1197. https://doi.org/10.1126/science.1168255.

(5) Bartels, L.; Meyer, G.; Rieder, K. H. Controlled Vertical Manipulation of Single CO Molecules with the Scanning Tunneling Microscope: A Route to Chemical Contrast. *Appl Phys Lett* **1998**, *71* (2), 213. https://doi.org/10.1063/1.119503.

(6) Meyer, G.; Zöphel, S.; Rieder, K. H. Controlled Manipulation of Ethen Molecules and Lead Atoms on Cu(211) with a Low Temperature Scanning Tunneling Microscope. *Appl Phys Lett* **1998**, *69* (21), 3185. https://doi.org/10.1063/1.117955.

(7) Goronzy, D. P.; Ebrahimi, M.; Rosei, F.; Arramel; Fang, Y.; Feyter, S. De; Tait, S. L.; Wang, C.; Beton, P. H.; Wee, A. T. S.; Weiss, P. S.; Perepichka, D. F. Supramolecular Assemblies on Surfaces: Nanopatterning, Functionality, and Reactivity. *ACS Nano* **2018**, *12*, 7445–7481. https://doi.org/10.1021/acsnano.8b03513.

(8) Barth, J. V; Costantini, G.; Kern, K. Engineering Atomic and Molecular Nanostructures at Surfaces. *Nature 2005 437:7059* **2005**, *437* (7059), 671–679. https://doi.org/10.1038/nature04166.

(9) Bartels, L. Tailoring Molecular Layers at Metal Surfaces. *Nature Chemistry 2010 2:2* **2010**, *2* (2), 87–95. https://doi.org/10.1038/nchem.517.

(10) Barth, J. V. Molecular Architectonic on Metal Surfaces. *Annu Rev Phys Chem* **2007**, *58*, 375–407. https://doi.org/10.1146/annurev.physchem.56.092503.141259.

(11) Simpson, G. J.; García-López, V.; Petermeier, P.; Grill, L.; Tour, J. M. How to Build and Race a Fast Nanocar. *Nature Nanotechnology 2017 12:7* **2017**, *12* (7), 604–606. https://doi.org/10.1038/nnano.2017.137.

(12) Simpson, G. J.; García-López, V.; Boese, A. D.; Tour, J. M.; Grill, L. How to Control Single-Molecule Rotation. *Nature Communications 2019 10:1* **2019**, *10* (1), 1–6. https://doi.org/10.1038/s41467-019-12605-8.

(13) Simpson, G. J.; García-López, V.; Boese, A. D.; Tour, J. M.; Grill, L. Directing and Understanding the Translation of a Single Molecule Dipole. *Journal of Physical Chemistry Letters* **2023**, *14* (10), 2487–2492.

https://doi.org/10.1021/ACS.JPCLETT.2C03472/ASSET/IMAGES/LARGE/JZ2C03472_0004.JPEG.

(14) Celotta, R. J.; Balakirsky, S. B.; Fein, A. P.; Hess, F. M.; Rutter, G. M.; Stroscio, J. A. Autonomous Assembly of Atomically Perfect Nanostructures Using a Scanning Tunneling Microscope. *Review of Scientific Instruments* **2014**, *85* (12), 121301. https://doi.org/10.1063/1.4902536.

(15) Meyer, G.; Moresco, F.; Hla, S. W.; Repp, J.; Braun, K. F.; Fölsch, S.; Rieder, K. H. Manipulation of Atoms and Molecules with the Low-Temperature Scanning Tunneling Microscope. *Japanese Journal of Applied Physics, Part 1: Regular Papers and Short Notes and Review Papers* **2001**, *40* (6 B), 4409–4413. https://doi.org/10.1143/JJAP.40.4409/XML.

(16) Eigler, D. M.; Schweizer, E. K. Positioning Single Atoms with a Scanning Tunnelling Microscope. *Nature 1990 344:6266* **1990**, *344* (6266), 524–526. https://doi.org/10.1038/344524a0.

(17) Nilius, N.; Wallis, T. M.; Ho, W. Tailoring Electronic Properties of Atomic Chains Assembled by STM. *Appl Phys A Mater Sci Process* **2005**, *80* (5), 951–956. https://doi.org/10.1007/S00339-004-3121-0/METRICS.

(18) Mokaberi, B.; Yun, J.; Wang, M.; Requicha, A. A. G. Automated Nanomanipulation with Atomic Force Microscopes. *Proc IEEE Int Conf Robot Autom* **2007**, 1406–1412. https://doi.org/10.1109/ROBOT.2007.363181.

(19) Crommie, M. F.; Lutz, C. P.; Eigler, D. M. Confinement of Electrons to Quantum Corrals on a Metal Surface. *Science (1979)* **1993**, *262* (5131), 218–220. https://doi.org/10.1126/SCIENCE.262.5131.218.

(20) Khajetoorians, A. A.; Wegner, D.; Otte, A. F.; Swart, I. Creating Designer Quantum States of Matter Atom-by-Atom. *Nature Reviews Physics 2019 1:12* **2019**, *1* (12), 703–715. https://doi.org/10.1038/s42254-019-0108-5.

(21) Gomes, K. K.; Mar, W.; Ko, W.; Guinea, F.; Manoharan, H. C. Designer Dirac Fermions and Topological Phases in Molecular Graphene. *Nature 2012 483:7389* **2012**, *483* (7389), 306–310. https://doi.org/10.1038/nature10941.

(22) Río, E. C.; Mallet, P.; González-Herrero, H.; Lado, J. L.; Fernández-Rossier, J.; Gómez-Rodríguez, J. M.; Veuillen, J. Y.; Brihuega, I. Quantum Confinement of Dirac Quasiparticles in Graphene Patterned with Sub-Nanometer Precision. *Advanced Materials* **2020**, *32* (30), 2001119. https://doi.org/10.1002/ADMA.202001119.

(23) Gutiérrez, C.; Walkup, D.; Ghahari, F.; Lewandowski, C.; Rodriguez-Nieva, J. F.; Watanabe, K.; Taniguchi, T.; Levitov, L. S.; Zhitenev, N. B.; Stroscio, J. A. Interaction-Driven Quantum Hall Wedding Cake–like Structures in Graphene Quantum Dots. *Science (1979)* **2018**, *361* (6404), 789–794. https://doi.org/10.1126/science.aar2014.

(24) Khajetoorians, A. A.; Wiebe, J.; Chilian, B.; Wiesendanger, R. Realizing All-Spin-Based Logic Operations Atom by Atom. *Science (1979)* **2011**, *332* (6033), 1062–1064. https://doi.org/10.1126/science.12017.

(25) Huff, T.; Labidi, H.; Rashidi, M.; Livadaru, L.; Dienel, T.; Achal, R.; Vine, W.; Pitters, J.; Wolkow, R. A. Binary Atomic Silicon Logic. *Nature Electronics 2018 1:12* **2018**, *1* (12), 636–643. https://doi.org/10.1038/s41928-018-0180-3.

(26) Eigler, D. M.; Lutz, C. P.; Rudge, W. E. An Atomic Switch Realized with the Scanning Tunnelling Microscope. *Nature 1991 352:6336* **1991**, *352* (6336), 600–603. https://doi.org/10.1038/352600a0.

(27) Wurman, P. R.; Barrett, S.; Kawamoto, K.; MacGlashan, J.; Subramanian, K.; Walsh, T. J.; Capobianco, R.; Devlic, A.; Eckert, F.; Fuchs, F.; Gilpin, L.; Khandelwal, P.; Kompella, V.; Lin, H.; MacAlpine, P.; Oller, D.; Seno, T.; Sherstan, C.; Thomure, M. D.; Aghabozorgi, H.; Barrett, L.; Douglas, R.; Whitehead, D.; Dürr, P.; Stone, P.; Spranger, M.; Kitano, H. Outracing Champion Gran Turismo Drivers with Deep Reinforcement Learning. *Nature* **2022**, *602*. https://doi.org/10.1038/s41586-021-04357-7.

(28) Kaufmann, E.; Bauersfeld, L.; Loquercio, A.; Müller, M.; Koltun, V.; Scaramuzza, D. Champion-Level Drone Racing Using Deep Reinforcement Learning. | *Nature* | **2023**, *620*. https://doi.org/10.1038/s41586-023-06419-4.

(29) Silver, D.; Huang, A.; Maddison, C. J.; Guez, A.; Sifre, L.; Van Den Driessche, G.; Schrittwieser, J.; Antonoglou, I.; Panneershelvam, V.; Lanctot, M.; Dieleman, S.; Grewe, D.; Nham, J.; Kalchbrenner, N.; Sutskever, I.; Lillicrap, T.; Leach, M.; Kavukcuoglu, K.; Graepel, T.; Hassabis, D. Mastering the Game of Go with Deep Neural Networks and

Tree Search. *Nature 2016 529:7587* **2016**, *529* (7587), 484–489. https://doi.org/10.1038/nature16961.

(30) Vinyals, O.; Babuschkin, I.; Czarnecki, W. M.; Mathieu, M.; Dudzik, A.; Chung, J.; Choi, D. H.; Powell, R.; Ewalds, T.; Georgiev, P.; Oh, J.; Horgan, D.; Kroiss, M.; Danihelka, I.; Huang, A.; Sifre, L.; Cai, T.; Agapiou, J. P.; Jaderberg, M.; Vezhnevets, A. S.; Leblond, R.; Pohlen, T.; Dalibard, V.; Budden, D.; Sulsky, Y.; Molloy, J.; Paine, T. L.; Gulcehre, C.; Wang, Z.; Pfaff, T.; Wu, Y.; Ring, R.; Yogatama, D.; Wünsch, D.; McKinney, K.; Smith, O.; Schaul, T.; Lillicrap, T.; Kavukcuoglu, K.; Hassabis, D.; Apps, C.; Silver, D. Grandmaster Level in StarCraft II Using Multi-Agent Reinforcement Learning. *Nature* **2019**, *575*. https://doi.org/10.1038/s41586-019-1724-z.

(31) Berner, C.; Brockman, G.; Chan, B.; Cheung, V.; Dennison, C.; Farhi, D.; Fischer, Q.; Hashme, S.; Hesse, C.; Józefowicz, R.; Gray, S.; Olsson, C.; Pachocki, J.; Petrov, M.; de Oliveira Pinto, H. P.; Raiman, J.; Salimans, T.; Schlatter, J.; Schneider, J.; Sidor, S.; Sutskever, I.; Tang, J.; Wolski, F.; Zhang, S. Dota 2 with Large Scale Deep Reinforcement Learning. 2021. https://doi.org/10.48550/arXiv.1912.06680.

(32) Novati, G.; de Laroussilhe, H. L.; Koumoutsakos, P. Automating Turbulence Modelling by Multi-Agent Reinforcement Learning. *Nature Machine Intelligence 2021 3:1* **2021**, *3* (1), 87–96. https://doi.org/10.1038/s42256-020-00272-0.

(33) Jesse, S.; Hudak, B. M.; Zarkadoula, E.; al -; Balke, N.; Morozovska, A.; Kalnaus, S.; Guo, S.; Bei, H. Exploring Electron Beam Induced Atomic Assembly via Reinforcement Learning in a Molecular Dynamics Environment. *Nanotechnology* **2022**, *33*, 115301. https://doi.org/10.1088/1361-6528/ac394a.

(34) Scheidt, J.; Diener, A.; Maiworm, M.; Müller, K. R.; Findeisen, R.; Driessens, K.; Tautz, F. S.; Wagner, C. Concept for the Real-Time Monitoring of Molecular Configurations during Manipulation with a Scanning Probe Microscope. *Journal of Physical Chemistry C* **2023**, 127–13817. https://doi.org/10.1021/acs.jpcc.3c02072.

(35) Shin, D.; Kim, Y.; Oh, C.; An, H.; Park, J.; Kim, J.; Lee, J. Deep Reinforcement Learning-Designed Radiofrequency Waveform in MRI. *Nat Mach Intell* **2021**, *3*, 985–994. https://doi.org/10.1038/s42256-021-00411-1.

(36) Krull, A.; Hirsch, P.; Rother, C.; Schiffrin, A.; Krull, C. Artificial-Intelligence-Driven Scanning Probe Microscopy. *Commun Phys* **2020**, *3* (1). https://doi.org/10.1038/S42005-020-0317-3.

(37) Kalinin, S. V; Ziatdinov, M.; Hinkle, J.; Jesse, S.; Ghosh, A.; Kelley, K. P.; Lupini, A. R.; Sumpter, B. G.; Vasudevan, R. K. Automated and Autonomous Experiments in Electron and Scanning Probe Microscopy. *ACS Nano* **2021**, *15* (8), 12604–12627. https://doi.org/10.1021/acsnano.1c02104.

(38) Wu, S.; Bai, H.; Jin, F. Automated Manipulation of Flexible Nanowires with an Atomic Force Microscope. *Conference Program Digest - 7th International Conference on Manipulation, Manufacturing and Measurement on the Nanoscale, IEEE 3M-NANO 2017* **2018**, *2018-January*, 229–235. https://doi.org/10.1109/3M-NANO.2017.8286320.

(39) Leinen, P.; Esders, M.; Schütt, K. T.; Wagner, C.; Müller, K. R.; Tautz, F. S. Autonomous Robotic Nanofabrication with Reinforcement Learning. *Sci Adv* **2020**, *6* (36). https://doi.org/10.1126/sciadv.abb6987.

(40) Chen, I.-J.; Aapro, M.; Kipnis, A.; Ilin, A.; Liljeroth, P.; Foster, A. S. Precise Atom Manipulation through Deep Reinforcement Learning. *Nature Communications 2022 13:1* **2022**, *13* (1), 1–8. https://doi.org/10.1038/s41467-022-35149-w.

(41) Powell, W. B.; George, A.; Bouzaiene-Ayari, B.; Simao, H. P. Approximate Dynamic Programming for High Dimensional Resource Allocation Problems. *Proceedings of the International Joint Conference on Neural Networks* **2005**, *5*, 2989–2994. https://doi.org/10.1109/IJCNN.2005.1556401.

(42) Ramsauer, B.; Simpson, G. J.; Cartus, J. J.; Jeindl, A.; García-López, V.; Tour, J. M.; Grill, L.; Hofmann, O. T. Autonomous Single-Molecule Manipulation Based on Reinforcement Learning. *Journal of Physical Chemistry A* **2023**, *127* (8), 2041–2050. https://doi.org/10.1021/acs.jpca.2c08696.

(43) Watkins, C. J. C. H.; Dayan, P. Q-Learning. 1992, pp 279–292.

(44) Sutton, R. S.; Barto, A. G. Reinforcement Learning: An Introduction, Second Edition. *The MIT Press* **2018**, 1–3.

(45) Russell, S. J.; Norvig, P. Artificial Intelligence: A Modern Approach Third Edition.

(46) Howard, R. A. *Dynamic Programming and Markov Processes*; The MIT Press and John Wiley & Sons, 1960.

TOC Graphic

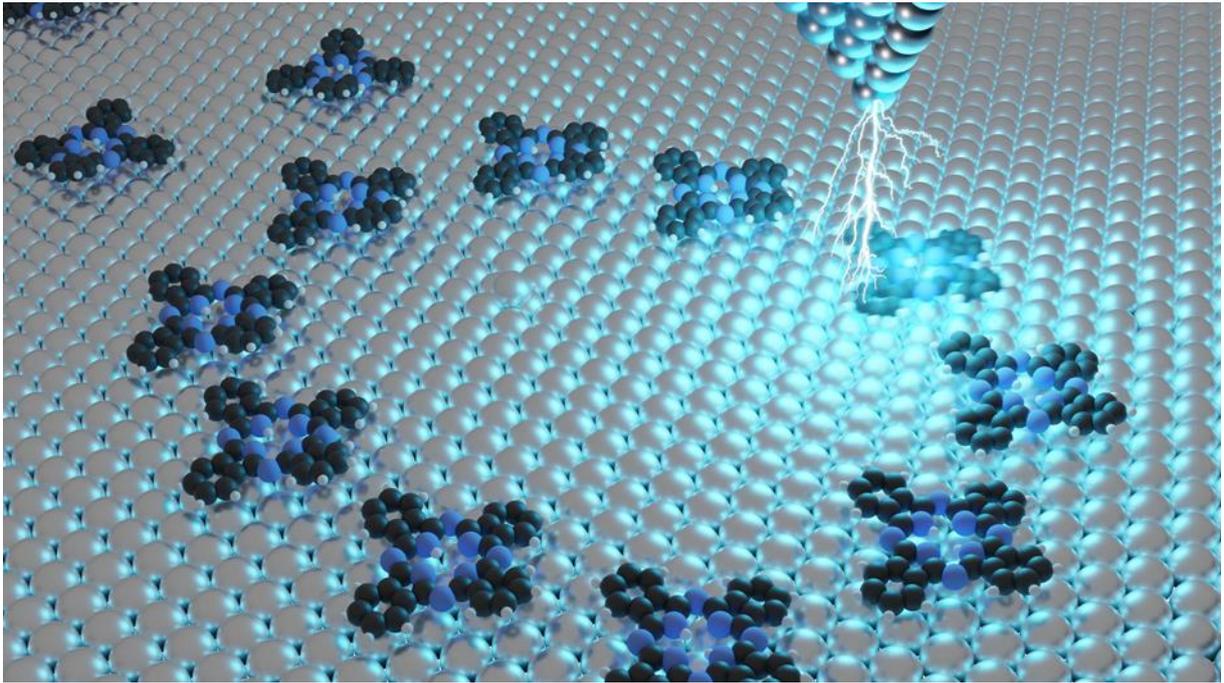